\documentclass[journal,12pt,onecolumn,draftclsnofoot,]{IEEEtran}

\usepackage[colorlinks]{hyperref}

\usepackage[cmex10]{amsmath}
\usepackage[utf8]{inputenc}
\usepackage{multirow}
\usepackage{soul}
\usepackage{algorithm}
\usepackage{algpseudocode}
\usepackage{graphicx}
\usepackage{dsfont}
\usepackage{xcolor}
\usepackage{cite}

\usepackage[utf8]{inputenc}
\usepackage{mathtools}
\usepackage[mathscr]{euscript}
\usepackage{amsmath}
\usepackage{amssymb}
\usepackage{amsfonts}
\usepackage{amssymb}
\usepackage{amsmath}
\usepackage{verbatim}
\usepackage[T1]{fontenc}
\usepackage[utf8]{inputenc}
\usepackage{latexsym}
\usepackage{enumerate}
\usepackage{indentfirst}
\usepackage{calligra}
\usepackage{ulem}
\usepackage{graphicx}
\usepackage{array}
\usepackage{float}
\usepackage{bbm}

\usepackage{mleftright}

\makeatletter
\let\old@ps@headings\ps@headings
\let\old@ps@IEEEtitlepagestyle\ps@IEEEtitlepagestyle
\def\psccfooter#1{%
    \def\ps@headings{%
        \old@ps@headings%
        \def\@oddfoot{\strut\hfill#1\hfill\strut}%
        \def\@evenfoot{\strut\hfill#1\hfill\strut}%
    }%
    \def\ps@IEEEtitlepagestyle{%
        \old@ps@IEEEtitlepagestyle%
        \def\@oddfoot{\strut\hfill#1\hfill\strut}%
        \def\@evenfoot{\strut\hfill#1\hfill\strut}%
    }%
    \ps@headings%
}
\makeatother

\usepackage{soul}

\usepackage{multirow}
\usepackage{soul,color}
\usepackage{graphicx}
\usepackage{dsfont}
\usepackage{subcaption}

\usepackage[utf8]{inputenc}
\usepackage{mathtools}
\usepackage[mathscr]{euscript}
\usepackage{amsmath}
\usepackage{amssymb}
\usepackage{amsfonts}
\usepackage{verbatim}
\usepackage[T1]{fontenc}
\usepackage[utf8]{inputenc}
\usepackage{latexsym}
\usepackage{enumerate}
\usepackage{indentfirst}
\usepackage{calligra}
\usepackage{ulem}

\usepackage{array}
\usepackage{float}
\usepackage{bbm}

\usepackage{geometry}
\usepackage{eurosym}

\usepackage{hyperref}

\usepackage{caption}

\usepackage{tikz}
\usepackage{pgfplots}

\pgfplotsset{compat=1.8}
\usepgfplotslibrary{statistics}
\pgfmathdeclarefunction{fpumod}{2}{%
        \pgfmathfloatdivide{#1}{#2}%
        \pgfmathfloatint{\pgfmathresult}%
        \pgfmathfloatmultiply{\pgfmathresult}{#2}%
        \pgfmathfloatsubtract{#1}{\pgfmathresult}%
        \pgfmathfloatifapproxequalrel{\pgfmathresult}{#2}{\def\pgfmathresult{5}}{}%
    }

\usepackage{tabularx}
\usepackage{flushend}
\usepackage{textcomp}
\usepackage{xcolor}
\usepackage{import}

\usetikzlibrary{trees,decorations,shadows}
\tikzset{level 1/.style={sibling angle=45,level distance=4mm}}
\usetikzlibrary{arrows.meta}
\usepackage{forest}
\usetikzlibrary{external}

\let\oldtikzexternalgetnextfilename\tikzexternalgetnextfilename \renewcommand{\tikzexternalgetnextfilename}[1]{\oldtikzexternalgetnextfilename{#1}\expandafter\tikzsetnextfilename\expandafter{#1}}

\usepackage{pgfplotstable}
\usepackage[outline]{contour}
\contourlength{1.2pt}

\pgfplotsset{compat=1.13} 

\usetikzlibrary{spy}
\usetikzlibrary{calc}
\usetikzlibrary{fadings}
\usetikzlibrary{patterns}
\usetikzlibrary{shadows}
\usetikzlibrary{mindmap}
\usetikzlibrary{backgrounds}
\usetikzlibrary{shapes.symbols}
\usetikzlibrary{shapes.multipart}
\usetikzlibrary{shapes.geometric}
\usetikzlibrary{automata,positioning}
\usetikzlibrary{decorations.fractals} 
\usetikzlibrary{decorations.markings}
\usetikzlibrary{decorations.pathreplacing}
\usetikzlibrary{decorations.pathmorphing}
    
\usepackage{bm}

\usepackage{nomencl}
\makenomenclature
\tikzset{edge from parent/.style={segment angle=10,draw}}

\tikzset{
  my rounded corners/.append style={rounded corners=2pt},
}

\def\BibTeX{{\rm B\kern-.05em{\sc i\kern-.025em b}\kern-.08em
    T\kern-.1667em\lower.7ex\hbox{E}\kern-.125emX}}

\renewcommand{\nomgroup}[1]{%
     \ifthenelse{\equal{#1}{O}}{\item[\textit{Operators}]}{%
        \ifthenelse{\equal{#1}{I}}{\item[\textit{Indices}]}{%
            \ifthenelse{\equal{#1}{A}}{\item[\textit{Acronyms}]}{%
            `\ifthenelse{\equal{#1}{V}}{\item[\textit{Variables and parameters}]}{}}}}}
\usepackage{scalerel}
\usetikzlibrary{svg.path}
\definecolor{orcidlogocol}{HTML}{A6CE39}
\tikzset{
    orcidlogo/.pic={
        \fill[orcidlogocol] svg{M256,128c0,70.7-57.3,128-128,128C57.3,256,0,198.7,0,128C0,57.3,57.3,0,128,0C198.7,0,256,57.3,256,128z};
        \fill[white] svg{M86.3,186.2H70.9V79.1h15.4v48.4V186.2z}
        svg{M108.9,79.1h41.6c39.6,0,57,28.3,57,53.6c0,27.5-21.5,53.6-56.8,53.6h-41.8V79.1z M124.3,172.4h24.5c34.9,0,42.9-26.5,42.9-39.7c0-21.5-13.7-39.7-43.7-39.7h-23.7V172.4z}
        svg{M88.7,56.8c0,5.5-4.5,10.1-10.1,10.1c-5.6,0-10.1-4.6-10.1-10.1c0-5.6,4.5-10.1,10.1-10.1C84.2,46.7,88.7,51.3,88.7,56.8z};
    }
}

\newcommand\orcidicon[1]{\href{https://orcid.org/#1}{\mbox{\scalerel*{ \begin{tikzpicture}[yscale=-1,transform shape]
                \pic{orcidlogo};
                \end{tikzpicture}
            }{|}}}}

\begin{document}
%
\title{Chance constrained day-ahead robust flexibility 
needs assessment
for low voltage distribution network}

\author{Md~Umar~Hashmi,~\IEEEmembership{Member,~IEEE}~\orcidicon{0000-0002-0193-6703},~Arpan~Koirala,~\IEEEmembership{Graduate~Student~Member~IEEE}~\orcidicon{0000-0003-4826-7137},~Hakan~Ergun,~\IEEEmembership{Senior~Member,~IEEE}~\orcidicon{0000-0001-5171-1986}, and~Dirk~Van~Hertem,~\IEEEmembership{Senior~Member,~IEEE}~\orcidicon{0000-0001-5461-8891}
\thanks{Corresponding author email: mdumar.hashmi@kuleuven.be}
\thanks{Md U. Hashmi, A. Koirala, H. Ergun,  and D. Van Hertem are with KU Leuven, division Electa \& EnergyVille, Genk, Belgium}
\thanks{This work is supported by the H2020 EUniversal project, grant ID: 864334 (\url{https://euniversal.eu/}) and the energy transition funds project BREGILAB organized by the FPS economy, S.M.E.s, Self-employed and Energy.}}

\maketitle


\begin{abstract}
For market-based procurement of low voltage (LV) flexibility, DSOs 
identify the amount of flexibility needed for resolving probable distribution network (DN) voltage and thermal congestion.
A framework is required to avoid over or under procurement of flexibility in the presence of uncertainty.
To this end, we propose a scenario-based robust chance-constrained (CC) day-ahead flexibility needs assessment (FNA) framework.
The CC level is analogous to the risk DSO is willing to take in flexibility planning.
Multi-period optimal power flow is performed to calculate the amount of flexibility needed to avoid network issues.
Flexibility is defined in terms of nodal power ramp-up and ramp-down and cumulative energy needs over a full day for each node. 
Future uncertainties are considered as multiple scenarios generated using multivariate Gaussian distribution and Cholesky decomposition.
These scenarios are utilized to solve the flexibility needs assessment optimal power flow (FNA-OPF) problem.
Zonal clustering of an LV feeder is performed using electrical distance as a measure and spatial partitioning. 
The FNA tool calculates ramp-up and ramp-down flexibility's power and energy requirements.
Energy and power needs are often valued differently in many energy markets. We identify the marginal value of flexibility associated with energy and power needs separately.
From numerical results for an LV feeder, it is observed that zonal flexibility needs assessment is more immune to uncertainty than nodal flexibility needs, making it more useful for DSOs to evaluate day-ahead flexibility procurement.
We also propose a Pareto optimal mechanism for selecting CC level to reduce flexibility needs while reducing DN congestion.
\end{abstract}


\maketitle

\tableofcontents

\pagebreak

\section{Introduction}
With growing uncertainty in distribution network (DN) operation due to distributed generation (DG) and loads with high simultaneity factor such as electric vehicle charging, it is crucial for DSOs to plan flexible resources to ensure reliable operation of DNs. 
ENTSO-e defines flexibility "as the ability of the power system
to cope with variability and uncertainty in demand, generation
and grid capacity, while maintaining a satisfactory level of
reliability at all times" \cite{entsoefact}.
Flexibility assists power networks to damp fluctuations, which could potentially reduce the reliability of the DN \cite{nosair2015flexibility, abdin2018integrated}. 
Authors in \cite{pinto2017multi} propose a computational method to calculate prosumer multi-period flexibility forecast. Using these feasible flexibility spaces, prosumers can participate in the energy market. Reference
\cite{ayon2017optimal} shows that small electricity prosumers can utilize their flexibilities for providing grid services while making a profit in day-ahead (DA) market. Authors in \cite{hooshmand2021optimal} schedule flexible resources for DA energy dispatch, which minimizes the energy procurement cost.
\cite{meibetaner2019co} presents a case study of Northern Germany, where every MW of additional wind installation will require 0.7 MW of flexibility for avoiding curtailment of renewable generation.
Prior works \cite{fonteijn2018flexibility,tsaousoglou2021mechanism,torbaghan2016local} present market mechanisms for end-user flexible resources for solving low voltage (LV) network issues.
Market mechanism presented in \cite{fonteijn2018flexibility} utilizes contractual flexibility by DSOs for H2020 Interflex project.
Authors in \cite{tsaousoglou2021mechanism} propose an efficient design for flexibility markets. 
Although flexibility market design is not the focus of the work, interested readers can refer to \cite{coninx2018gets}, \cite{zhang2018power}.
In \cite{agbonaye2021mapping} authors observe that locational temporal flexibility mapping is crucial for avoiding future network issues. They propose a mapping tool for Northern Ireland through GIS analysis.

Activation of flexibility in LV DN could aim at (a) mitigating LV DN issues \cite{arboleya2022flexibility} and/or (b) providing services to MV and HV transmission or DNs. The latter would require coordination in the operation of flexible resources, ensuring it does not create new problems \cite{givisiez2020review}.
The focus of this work is to utilize flexible resources for solving LV DN issues.
\textcolor{black}{Flexibility needs assessment (FNA) is fairly a new concern for power networks. Recent works such as \cite{davidov2020novel, laur2020optimal, hillberg2019flexibility,  yang2020flexibility} discuss the need for flexibility provisioning. \cite{davidov2020novel} utilize FNA for prioritization of future investments. Authors in \cite{laur2020optimal, yang2020flexibility} quantifies flexibility provisions based on conditional value at risk.}
\textcolor{black}{
The goal of this paper is to quantify the flexibility needs of a DN from a DSO's perspective. The DSO utilizes the FNA output provided by the framework proposed in this work for procurement of flexible resources in DA flexibility market implemented in H2020 EUniversal project.}
The procurement of such flexible resources in the DA market can be costly. 
Under-procuring can lead to network congestion, and over-procuring of flexible resources may not be efficient.
FNA should consider the probable uncertainties which could happen. 
Advanced time planning utilizes one or a combination of  (a) probabilistic modelling \cite{lu2018probabilistic}, (b) scenario generation \cite{chen2018model}, or (c) robust planning \cite{conejo2021robust, baringo2010offering, bertsimas2006robust}.
We propose a scenario-based robust DA planning mechanism for quantifying the flexibility needs of a DN. 
Chance constrained (CC) levels are utilized to avoid planning for the worst-case scenario.
{The DSO also needs to consider how a selection of a CC level would project on \textcolor{black}{probable} DN congestion.}

Power network clustering is widely utilized for operation and planning of transmission and DNs \cite{blumsack2009defining, sanchez2014hierarchical, ding2018clusters}.
In \cite{blumsack2009defining, sanchez2014hierarchical} network clustering is used for transmission system operator level resource dispatch and planning.
Authors in \cite{ding2018clusters} use zonal voltage regulation in the context of a LV DN with high penetration of DG.
In this work, we identify zones using spectral partitioning, along with $k$-means clustering for a doubly stochastic measure matrix formulated using a static network admittance matrix.
\textcolor{black}{The proposed framework for zonal clustering partitions the DN in connected segments, a key challenge in other partitioning methods.}
These zones are used to aggregate nodal flexibility needs.
Often, exact flexibility needs may not be met by the resources available in the flexibility market. In such a case, zonal FNA can be utilized by DSOs to find alternative resources which could resolve the network congestion.
Advantages of flexibility aggregation are also observed in \cite{torbaghan2019optimal}.

\textcolor{black}{In robust optimization, an uncertainty set is defined, over which a robust feasible solution is identified. It is unclear how we can define such an uncertainty set for FNA problem. Thus, we utilize generated scenarios for creating bounds on uncertainty. These scenarios define the uncertainty set for the flexibility needs assessment optimal power flow (FNA-OPF).
The bounded uncertainty and its associated worst-case action selection may lead to too conservative solution
(see Chapter 2 of \cite{ben2002robust}). 
A simple scenario generation model is proposed which takes a point forecast and its associated forecast error as input for generating nodal load and generation scenarios.}
\textcolor{black}{
The flexibility needs which provides a feasible solution for all scenarios will often lead to an over-procurement of flexible resources. These values of flexibility needs will be robust in true sense. To avoid worst case planning, chance-constrained is used. CC level is analogous to the risk level DSO is willing to consider in flexibility planning.
Chance constraint is associated to the FNA. 
The generated scenarios are used as snapshots of Monte Carlo simulations. 
}



FNA-OPF calculates the flexibility needs for solving LV DN issues while minimizing the cost of curtailed load and generation. 
Furthermore, spatio-temporal metrics of FNA are proposed.
Numerically, it is observed that variance of the day-ahead FNA can be reduced by zonal aggregation of flexibility needs. This implies aggregation makes it more predictable for the DSO to plan flexibility in a day-ahead setting. The zonal FNA can help DSOs to identify alternative flexible resources which could probably solve DN congestion issues.
{We also propose a Pareto optimal mechanism for selecting the level of CC used in FNA-OPF.}
{The selection of CC level for proposed FNA of a DN should consider DSO's risk for over and under-procurement of flexible resources. }
{In multi-objective optimization, some of the objectives may be in conflict with others. In such a case, an optimal solution for one objective could lead to unacceptable outcomes for other objectives. In power systems, prior works such as \cite{elattar2020optimal, li2018two, knezovic2017robust} utilize multi-objective optimization. Authors in \cite{knezovic2017robust} use epsilon-constrained Pareto front for an electric vehicle scheduling problem of an aggregator, while considering DSO's goal to minimize losses at the same time reducing the charging cost. In our work, we use the Pareto optimal value of chance-constraint to reduce DSO's risk of over-procurement of flexible resources while reducing the probable DN congestion incidents.}

\textcolor{black}{
The key contributions of the paper are:\\
$\bullet$ \textit{Uncertainty consideration}: 
We propose a simple scenario generation model using multivariate Gaussian distribution and Cholesky decomposition,  which takes as input PV size at a node, forecasted nodal load profiles, forecast of normalized PV generation,  and the associated forecast errors for load and solar generation. The forecast error creates a boundary of uncertainty. The generated scenarios serve as input for the FNA-OPF.\\
$\bullet$ \textit{Flexibility needs assessment}: 
The spatial and temporal robust flexibility needs of a DN are identified for avoiding all network incidents for all scenarios generated. For risk-averse FNA of the DN, a chance constrained nodal flexibility level is applied. Further, the nodal FNA is extended to zonal FNA by aggregating flexibility needs in a zone.  \\
$\bullet$ \textit{Case studies}: Two dedicated numerical case studies are presented. The first case study shows the FNA output of a DN. A Pareto optimal mechanism is provided for tuning CC level. The selection of CC level is projected on to unavoided network incidents using power flow simulations. The FNA of a DN is aggregated for providing zonal FNA. 
The second case study quantifies the marginal value of flexibility based on energy or power needs. It is observed that for the DN, power needs are twice as important as the energy needs.
}
\begin{figure}[!htbp]
	\center
	\includegraphics[width=5in]{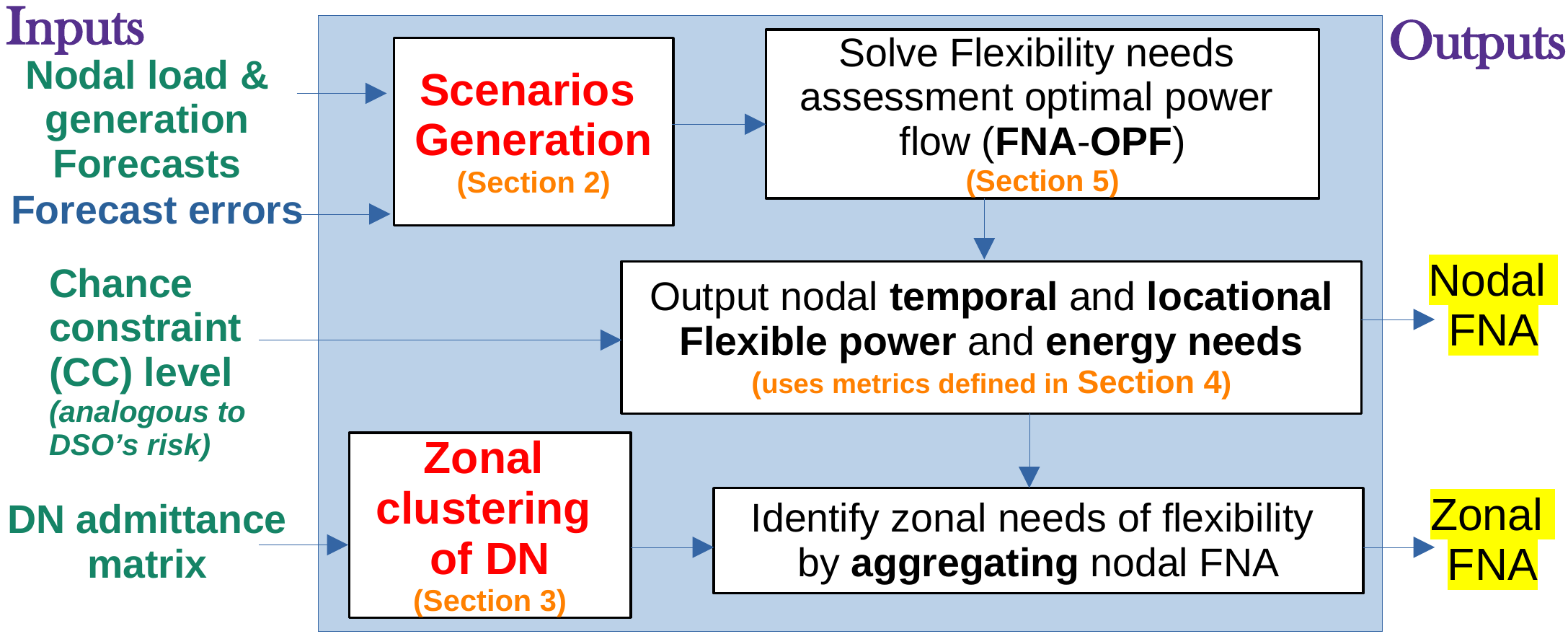}
	\vspace{-2pt}
	\caption{\small{Structure of this paper and the inputs and the outputs of robust chance-constrained FNA under uncertainty. }}
	\label{fig:flexassess}
\end{figure}

The structure of the paper and the inputs and outputs of the proposed FNA framework are shown in Fig.~\ref{fig:flexassess}.
In Section \ref{section2}, we detail the methodology to generate scenarios.
In Section \ref{section3}, zonal clustering of a DN based on electrical distance using spatial partitioning is provided.
In Section \ref{section4}, the metrics used in FNA are quantified.
In Section \ref{section5}, the FNA-OPF is detailed.
Section \ref{section6} presents the numerical results.
Section \ref{section7} concludes the paper.

\section{Scenario generation}
\label{section2}
\textcolor{black}{The focus of the present paper is to consider probable future uncertainties to provide FNA. This tool is developed as part of the EUniversal project to be implemented by the distribution system operator Mitnetz Storm in Germany. Due to German data limitations, only historical forecast errors and point forecasts are available for scenario generation. This has led us to opt for a simple scenario generation model with data limitation.}
The generated scenarios using 
multivariate Gaussian distribution and Cholesky decomposition 
considers (a) correlation of data points, (b) parameter distribution.
Cholesky factor is utilized to form unique scenarios
emulating different uncertainties.
This structure of scenario generation 
can be used for historical data and point forecasts with known forecast errors.
{In the present case,} large historical data is unavailable, instead a point forecast along with its associated forecast error is known. In such a case, a multivariate Gaussian distribution can be formed by using point time-series forecast as the mean value and the variance of the distribution is proportional to forecast error \cite{chakraborty2006generating}. 

\subsection{Multivariate distribution and Cholesky decomposition}
{The multivariate Gaussian distribution is a generalization of the one-dimensional normal distribution to more than one dimension.}
For our case, $X_i$ denotes the load profile of consumers connected at time $i$.
The multivariate normal distribution of a $S$-dimensional random vector $\mathbb{X} = (X_1, X_2, .... , X_S)^T$ can be written as
$
    \mathbb{X} \sim \textit{N} (\mu, \Sigma),$
with $S$-dimensional mean vector given as
$
    \mu =  \textbf{E}[\mathbb{X}] = 
    \mleft[ \begin{array}{c}
         \textbf{E}[\mathbb{X}_1] , ..,
         \textbf{E}[\mathbb{X}_S] 
    \end{array} \mright]^T, 
$
and the covariance matrix is given as
$
    \Sigma  = 
    \mleft[ \begin{array}{ccc}
         \texttt{Cov}(\mathbb{X}_1,\mathbb{X}_1)  & .. & \texttt{Cov}(\mathbb{X}_1,\mathbb{X}_S) \\
        : &  \ddots & :\\
        \texttt{Cov}(\mathbb{X}_S,\mathbb{X}_1)  & .. & \texttt{Cov}(\mathbb{X}_S,\mathbb{X}_S) \\
    \end{array} \mright], 
$
where $\texttt{Cov}(\mathbb{X}_i,\mathbb{X}_j)$ (also denoted as $\Sigma_{i,j}$) denote the covariance between $\mathbb{X}_i$ and $\mathbb{X}_j$ and is given as
\begin{equation}
    \Sigma_{i,j} = \textbf{E}[(\mathbb{X}_i - \mu_i)(\mathbb{X}_j - \mu_j)], ~i,j \in\{1,..,S\}.
\end{equation}
Each component $X_i$ has distribution $N(\mu_i, \sigma_i^2)$ with $\Sigma_{ii} = \sigma_i^2$.
The covariance matrix is a square matrix. 
To qualify as a covariance matrix, $\Sigma$ must be symmetric ($\Sigma = \Sigma^T$) and positive semidefinite ($x^T\Sigma x \geq 0 $). 


In order to ensure that Cholesky decomposition does not fail, a small error is added to the covariance matrix for numerical reasons, e.g. $
    \Sigma = \Sigma + \epsilon I_{S},
$
where $I_{S}$ denotes identity matrix of order $S$, $\epsilon$ denotes small error value.
This ensures that eigenvalues of $\Sigma$ do not decay rapidly, which stabilizes the decomposition. Due to the small magnitude of $\epsilon$, it has inconsequential effects on the samples while ensuring numerical stability \cite{ki2006gaussian,multivariate_blog}.

Cholesky decomposition is used to calculate the lower triangular matrix, $L$, for the covariance matrix such that
$
    \Sigma =  L L^T.
$
$L$ is also called the Cholesky factor.
A lower triangular matrix is convenient because it reduces the calculation of a scenario as $\mu + LZ$ where $Z \sim N(0,I)$ to 
\begin{gather*}
    X_1 = \mu_1 + l_{11} z_1,\\
    X_2 = \mu_2 + l_{21}z_1 + l_{22}z_2, ~\ldots\\
    X_S = \mu_S + l_{S1}z_1 + l_{S2}z_2 + \ldots +l_{SS}z_S
\end{gather*}
where $z_1, z_2, \ldots, z_S$ are
independent and identically distributed random variables.
The Cholesky factor is utilized to form unique scenarios \cite{muschinski2021cholesky} and is calculated using eigenvalue factorization. 
Since $\Sigma$ is symmetric and positive semidefinite, therefore, its eigenvalues $\lambda_1, \ldots, \lambda_S$ are non-negative. 
The covariance matrix can be denoted as $\Sigma = V \wedge V^T$, where $V$ is an orthogonal matrix, i.e. $VV^T =I$, $\wedge$ is a diagonal matrix with eigenvalues as diagonal entries. In this case, $L=V\wedge^{0.5}$.

Consider $J$ scenarios are generated for a time-series with $S$ intervals. The scenario matrix is given as
\begin{equation}
    S_J = 
    \mleft[ \begin{array}{ccc}
         \mu_1 & ..&\mu_1 \\
         \mu_2 & ..&\mu_2 \\
         : & ..&:\\
         \mu_S & ..&\mu_S
    \end{array} \mright]_{S,J} 
    + \mleft[ \begin{array}{ccc}
         l_{11}& .. & 0 \\
         l_{21}& .. & 0 \\
        :&  \ddots & :\\
        l_{S1}&.. & l_{SS} \\
    \end{array} \mright] 
    \texttt{rnd}(S,J),
    \label{eq:scenariosgen}
\end{equation}
where $\texttt{rnd}(S,J) \sim N(0,I)$ denotes uniformly distributed random numbers.
\eqref{eq:scenariosgen} is used to generate $J$ number of unique forecast profiles. 

\subsection{Temporal and spatially correlated scenarios}
\begin{figure}[!htbp]
	\center
	\includegraphics[width=5in]{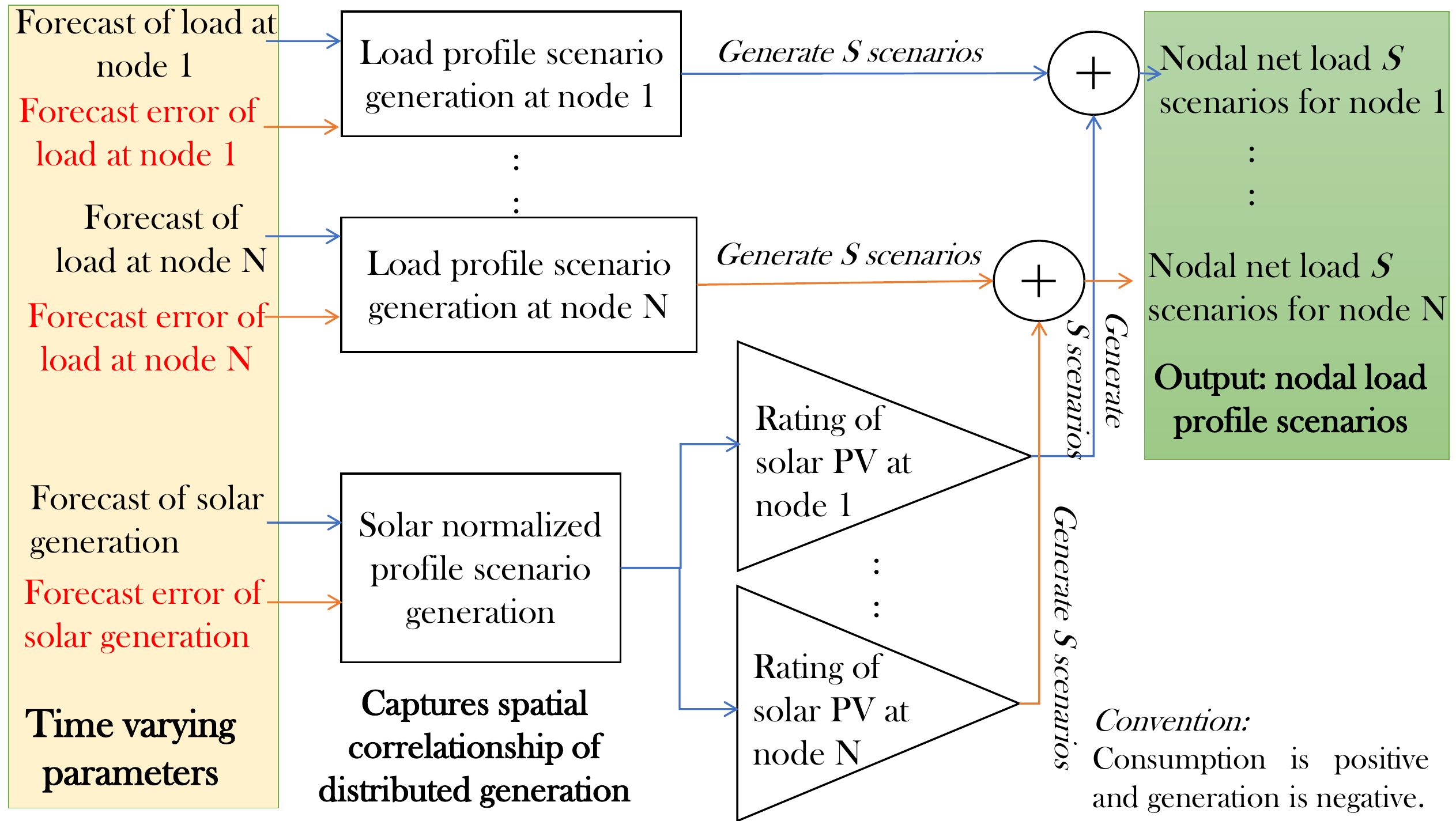}
	\vspace{-1pt}
	\caption{\small{Scenario generation decoupling load and generation to consider temporal and spatial correlationship. }}
	\label{fig:spatialtemporal}
\end{figure}

\textcolor{black}{The proposed scenario generation tool considers the point forecast as the mean shown in 
\eqref{eq:scenariosgen}.
The covariance matrix is a function of forecast error calculated based on historical data.
The forecast error is assumed to be 1.96 times the standard deviation of the normal distribution. This would ensure that more than 95\% of the incidents are covered. The extreme tail events are ignored in the scenario generation.}
\textcolor{black}{The scenario generation does not consider the spatial correlation. However, the generated scenarios do not ignore spatial correlation entirely.
The net load of a node consists of a solar generation as DG and load. Since consumption is assumed positive, therefore, the load is strictly non-negative and DG is negative. The load profile is sampled separately depending on load forecast error and its point forecast, both of which are assumed to be known.
On the other hand, DG scenarios are calculated based on its point forecast and associated forecast errors. The DG forecast scenarios are normalized in per kW installed capacity. These scenarios are assumed to be the same for all DGs in the feeder, assuming all DGs are installed in geographical proximity to be not affected by different levels of solar irradiance. 
The average spread of 160 Spanish feeders, \cite{koirala2020non}, is a square of size of 158 meters.
Therefore, this assumption would be valid for most DNs.
Fig. \ref{fig:spatialtemporal} shows input and output of scenario generation. 
}

\pagebreak

\section{Identifying zones of a LV DN}
\label{section3}
Identifying the zones of a LV DN will help the DSO in planning the flexibility needs of a network. Although, the analytical division of a DN is expected to work well, however, due to the sheer number of DN feeders, it becomes crucial to have a standardized framework for dividing DN into zones based on electrical and/or geographical distances.
For example, in the UK, there are more than 1.3 million LV feeders \cite{rigoni2015representative}.
In this section, we develop a clustering framework to identify the best-suited LV DN zonal partition using electrical distance as a measure.

\textcolor{black}{The proposed zone formation solves following challenges: (i)
The formation of connected zones requires an incidence matrix based measure, therefore, we consider admittance as a measure. (ii) The admittance matrix cannot be used directly, therefore, spectral decomposition of doubly stochastic matrix is used. (iii)  The adequate number of zones is not known a priori. However, to apply unsupervised clustering techniques such as \textit{k}-means, one should know how many clusters are needed. We use silhouette score as a measure for identifying the best number of clusters.\\
Although, the different methods mentioned above have been well discussed in the literature, in this work, they have been for the first time combined to perform zonal FNA of a DN.}
\subsection{Distance measure for clustering}
A DN can be represented in standard directed graph notation $G=(\mathscr{N},E, W)$, where $\mathscr{N}$ denotes the set of nodes such that $|\mathscr{N}| = D$ ($D$ denotes the total number of nodes), $E$ denotes edges and $W$ denotes edge weight matrix.
We aim to divide $\mathscr{N}$ into $p$ groups $\{M_1, M_2,...,M_p \}$ such that $M_i \subset \mathscr{N}$ and $M_i \cap M_j = \emptyset, ~ \forall i \neq j$.
The edge weights can be considered as a penalty for cutting that line in the graph while clustering.
It also measures the connection strength of two nodes in a graph.
A good measure would separate the weakly connected nodes and identify nodes that should be clustered together.

A power network can be clustered based on static network parameters such as line impedance/admittance or dynamic parameters such as power flows, line losses, voltage variations \cite{zhang2018power, sanchez2014hierarchical, cotilla2013multi}.
The electrical distance matrix is used for partitioning LV DNs in
\cite{blumsack2009defining, sanchez2014hierarchical}. Similar to these prior works, we use the line admittance matrix as the distance measure parameter. The weight $w_{ij} = Y_{ij} = 1/|R_{ij} + jX_{ij}|$, where $Y_{ij}$, $R_{ij}$ and $X_{ij}$ denotes admittance, resistance and impedance between nodes $i$ and $j$ and the edge weight $w: \mathscr{N}\times \mathscr{N} \rightarrow \mathbb{R}^{\geq 0}$ such that (a) $w_{ij} = w_{ji}, ~ \forall i, j$, (b) $w_{ij} = 0, \text{ if } (i,j) \notin E$, (c) $w_{ii} = 0, \forall i$.

\subsection{Spectral partitioning}
Spectral clustering is used for power network partitioning or creation of zones or network reduction in \cite{ding2018clusters, sanchez2014hierarchical}.
Previously described weight matrix
cannot be directly used, as the diagonal elements are all zeros. Authors in \cite{sanchez2014hierarchical} use the normalized Laplacian matrix for spectral partitioning.
Tutorial \cite{hespanha2004efficient} transforms the weight matrix into a doubly stochastic matrix. 
A double stochastic matrix is a special type of Markov matrix where not only each row but also each column add to 1.
Spectral properties of doubly stochastic matrices have been detailed in \cite{mourad2012spectral}.
For this transformed matrix, all eigenvalues are real and smaller than or equal to 1, with one eigenvalue exactly equal to 1.
For identifying $k$ partitions in a graph, the $k$ highest eigenvalues and corresponding orthonormal 
eigenvectors are identified.
The eigenvector matrix of the order $N \times k$ is used for DN partitioning, in effect reduces the dimensionality of the problem.
$k$-means clustering is used to partition the spectral data points.

\subsection{Goodness of a cluster}
Previously, we detailed partitioning a given LV DN into $k$ zones. In this subsection, we deal with identifying the best-suited values of $k$. In a real-world LV DN network partitioning problem, we may not know how many clusters we want. We use performance indices for measuring the goodness of a partition and, based on different values of $k$, identify the best value which fits our needs.
In this work, the goodness of a cluster is measured using the mean silhouette index of the network cluster.
The silhouette coefficient of a node is a confidence indicator of its association in a group $M_x$ \cite{ding2018clusters, scarlatache2012using}.
Rousseeuw \cite{rousseeuw1987silhouettes} proposed an interpretation based on value of silhouette coefficient. 
The zone selection algorithm is detailed in Appendix~\ref{zoneselection}.




\pagebreak

\section{Day-ahead flexibility planning}
\label{section4}
Flexibility is referred to as resources that can be activated or deactivated, thus in effect increasing or decreasing the consumption based on the grid's needs.
For a DN, nodal voltages and line loadings should remain within operational bounds. Violation of these limits could damage appliances, affect DN operation, or grid elements.
Authors in \cite{coninx2018gets} present two business models for operating electrical flexibility in the context of services provided to transmission and distribution system operators and assisting them to accommodate a greater amount of renewable energy. 
Depending on the application of flexibility, flexible resources needs can be measured in terms of response speed (or ramp rate), duration (or energy) and direction (ramp up and ramp down power) \cite{tang2021energy}.
Next, the framework 
for calculating the flexibility needs that DSO procures in the energy market is detailed.

\subsection{Flexibility: definition}
Prior work, \cite{ulbig2015analyzing}, used three parameters for defining flexibility: (a) ramp rate, (b) power, and (c) energy.
The flexibility model used in this work utilizes the power and energy parameters. This model resembles the simple storage model used in \cite{hashmi2020storage}.
The active power flexibility at node $i$ and time $t$ is 
given as
\begin{equation}
    \Delta P_{i,t}^{\text{flex}} = \Delta P_{i,t}^{\text{flex}+} + \Delta P_{i,t}^{\text{flex}-},
    \label{eq:flexdef}
\end{equation}
where $\Delta P_{i,t}^{\text{flex}+}$ and $\Delta P_{i,t}^{\text{flex}-}$ denotes ramp down and ramp up flexibility, respectively.

Ramp down flexibility decreases the nodal load and is analogous to load curtailment.
Ramp up flexibility increases the nodal load and is analogous to generation curtailment.
Ramp up flexibility can be provided by consumer loads such as HVAC, water heaters, pool pumps \cite{chen2018distributed} etc.
Ramp down flexibility can be provided by solar generation, {and also by curtailment of flexible loads}.
Prosumer energy storage if not fully charged or discharged can provide both ramp up and ramp down flexibility \cite{buvsic2017distributed}.
Note these flexible resources have a limited ramp rate, power and energy to be utilized as flexibility. As an example, the temperature of a water heater can be temporally constrained to be within some bounds of water temperature. If utilized as ramp up flexibility to increase the nodal load, it can only be operated till water temperature hits the upper bound (or the lower bound for ramping down). 
Thus, to respect technical device and user comfort constraints, we use power and energy constraints to define flexible resources.
The ramp down and ramp up flexibility energy for node $i$ are given as
\begin{equation}
    \begin{split}
        E_{i}^{\text{flex}+} = \sum_t \Delta P_{i,t}^{\text{flex}+} \Delta t,~~
        E_{i}^{\text{flex}-} = \sum_t \Delta P_{i,t}^{\text{flex}-} \Delta t,
    \end{split}
\end{equation}
with
\begin{gather}
\label{eq:flexrange1}
    \Delta P_{i,t}^{\text{flex}+} \in [0, \Delta P_{i,t, \max}^{\text{flex}+}], ~~
    \Delta P_{i,t}^{\text{flex}-} \in [\Delta P_{i,t, \min}^{\text{flex}-}, 0], ~ \forall t,\\
    \label{eq:flexrange3}
    E_{i}^{\text{flex}+} = [0, E_{i, \max}^{\text{flex}+}],~~~~
        E_{i}^{\text{flex}-} = [E_{i, \min}^{\text{flex}-}, 0].
\end{gather}

Since ramp up and ramp down flexibility levels can compensate each other, therefore, they are dealt with separately in order to correctly assess the DN needs.
Next, we describe the metrics used to quantify flexibility needs.

\subsection{Metrics for day-ahead FNA}
The limits on ramp up and down power and energy needs are infrastructure constraints that the optimization needs to take into account. 
The needs assessment should provide the temporal and locational needs of the DN.
With these inputs and dimensions to the needs assessment problem, the following metrics are utilized:
\begin{enumerate}
    \item \textit{Objective function}: the optimization minimizes the cost of operation of flexible resources while ensuring network issues are avoided. Assuming ample amount of flexible resource available, there are many solutions that can solve DN issues, some of which could have the same objective function value. 
    In this work, we utilize a flat penalty factor for penalizing the use of ramp down and ramp up flexibilities. The penalty factors are denoted as $\lambda^{\text{rampDown}}$, $\lambda^{\text{rampUp}}$. These factors are fixed for all nodes across all times.
    \item \textit{Flexibility power needs assessment metrics}: Ramp up ($R_{i,t}^{P_{\text{up}}})$ and ramp down ($R_{i,t}^{P_{\text{down}}})$ power (with CC) needs for node or zone $i$ and time $t$: the temporal distribution of ramp up and ramp down power needs will often have a fat-tailed distribution, as shown in Fig. \ref{fig:flexnee}. Planning for the absolute worst case may lead to drastic over-design of flexibility needs. 
    \item \textit{Flexibility energy needs assessment metrics}: Ramp up ($R_{i}^{E_{\text{up}}})$ and ramp down ($R_{i}^{E_{\text{down}}})$ energy with chance constrained needs for node or zone $i$ denotes the energy needs of DN. 
    
    \begin{figure}[!htbp]
	\center
	\includegraphics[width=4.5in]{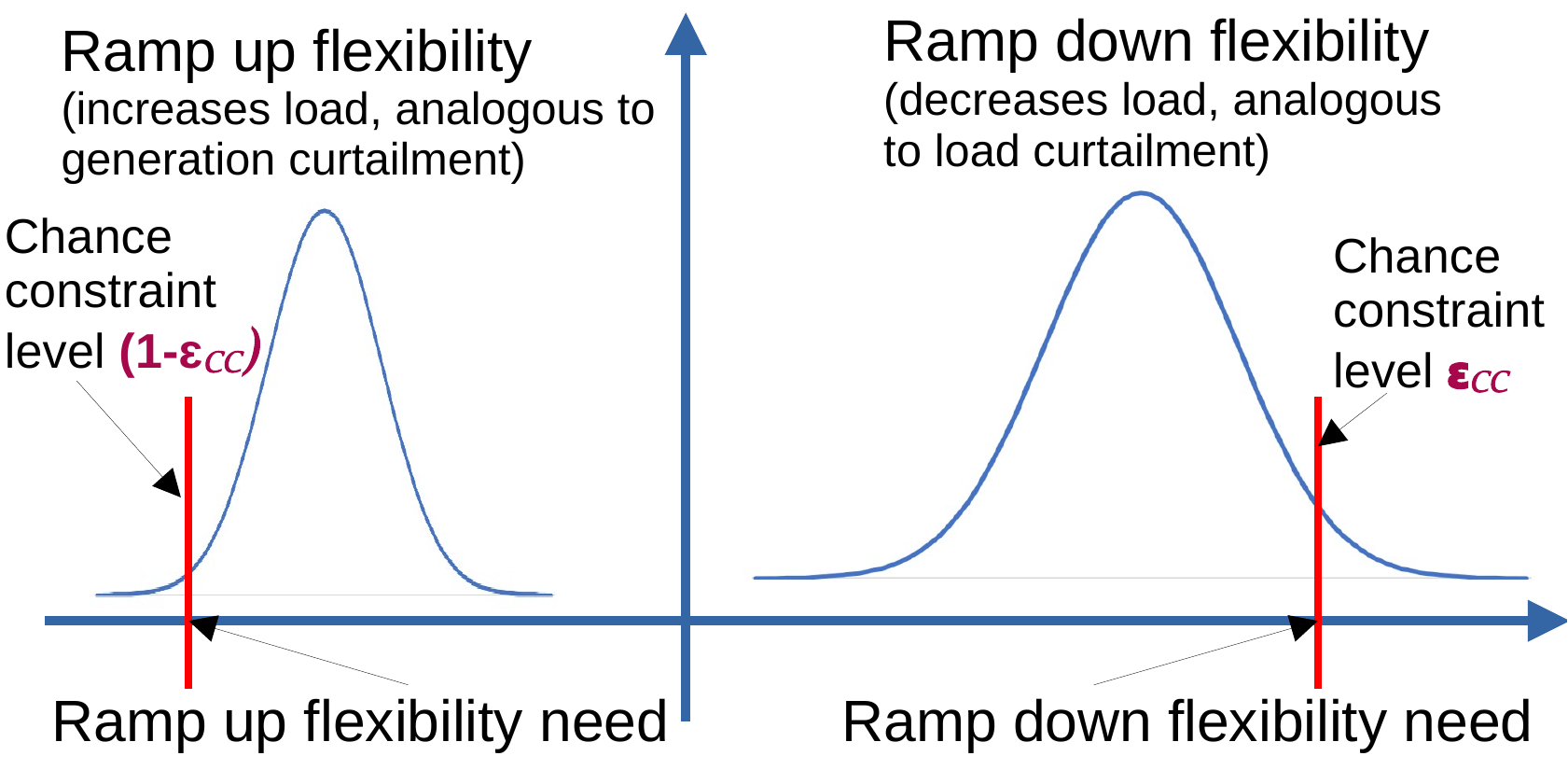}
	\caption{\small{FNA metrics for ramp up and down power and energy using chance constraint level denoted as $\epsilon_{cc}$}}
	\label{fig:flexnee}
\end{figure}
    \item The nodal and zonal FNA are compared using the average weighted variance of nodal ramp up and down power and energy needs for all $t$.
\item {
\textit{Tuning CC level ($\epsilon_{cc}$) for FNA}: The selection of $\epsilon_{cc}$ is crucial for DSO to appropriately plan for day-ahead flexibilities. Both over and under procurement is not efficient. We project $\epsilon_{cc}$ on mean DN voltage and thermal congestion over all generated scenarios. Next, the Pareto optimal level of CC is utilized for reducing DN flexibility activation and reducing DN congestion events. Pareto optimality is frequently used to select between conflicting goals of a multi-objective optimization problem \cite{deb2005searching}.
}
\end{enumerate}









\pagebreak

\section{Flexibility needs assessment}
\label{section5}
\subsection{Notation}


A power network consists of nodes ($\mathscr{N}$),
branches ($E$), generators and loads. 
Each node $i \in \mathscr{N}$ has two variables, i.e., voltage magnitude ($V_{i,t}$) and phase angle ($\theta_{i,t}$) for time $i$ which depends on power injection and load magnitude. The branch admittance $(i,j) \in E$ together with the nodal voltage differences govern the power flow and losses.
The subset of nodes with loads connected is denoted as $\mathscr{N}_L \subset \mathscr{N}$. The active and reactive power demand in these nodes are denoted as $P^d_{i,t}$ and $Q^d_{i,t}$.
Nodes with generators connected is denoted as $\mathscr{N}_G \subset \mathscr{N}$, have active and reactive power generation denoted as $P^g_{i,t}$ and $Q^g_{i,t}$.

\subsection{Optimization formulation}
The flexibility needs assessment is performed by solving the optimal power flow problem which minimizes the cost of dispatching flexible resources. 
We solve a multi-period optimal power flow problem with penalties associated with generation and load curtailment, refer to \eqref{eq:flexrange1}. Further, multi-period coupled energy curtailment constraint is considered along with instantaneous curtailed power constraint, refer to  \eqref{eq:flexrange3}. The optimization problem is denoted as flexibility needs assessment optimal power flow dispatch or FNA-OPF. FNA-OPF is an adaptation of the resource dispatch tool introduced in \cite{hashmi2021sest}.
\begin{subequations}
\label{eq:fnaoptimization}
\begin{equation}
\label{eq:porgobjective}
   ({\text{FNA-OPF}})~\underset{\substack{\Delta P^{\text{flex}}_{i,t}}}{\text{min}}~ \sum_i \Big( \lambda^{\text{rampDown}}  E_i^{\text{flex}+} - 
   \lambda^{\text{rampUp}} E_i^{\text{flex}-} \Big)
\end{equation} 
\text{subject to,~~\eqref{eq:flexrange1},  \eqref{eq:flexrange3}} 
\begin{equation}
   ~V_{\min}^i \leq |V_{i,t}| \leq V_{\max}^i, ~ \forall i \in \mathscr{N} , t \in \{1,..,T\},
  \label{eq:voltage}
\end{equation}
\begin{equation}
 (P^g_{i,t}) - (P^d_{i,t} - \Delta P^{\text{flex}}_{i,t}) + jQ^d_{i,t} = \sum_j s_{i,j,t},  ~~\forall~ i,j \in \mathscr{N} ,
 \label{eq:powerbalance}
\end{equation}
\begin{equation}
  |s_{i,j,t}| \leq s_{i,j}^{\max}, ~~\forall~ i,j \in \mathscr{N}     \label{const:thermal}, 
\end{equation}
\begin{equation}
  P^g_{i,t} \in [P^g_{\min,i},P^g_{\max,i}] , ~~\forall~ i \in \mathscr{N_G}     \label{const:genlimit}, 
\end{equation}
\begin{equation}
  s_{i,j,t}  = \textbf{Y}_{ij}^*V_{i,t}V_{i,t}^* - \textbf{Y}_{ij}^*V_{i,t}V_{j,t}^*, ~ \forall (i,j) \in E \cup E^R, 
  \label{eq:ohmslaw}
\end{equation}
\begin{equation}
  \angle (V_{i,t} V_{j,t}^*) \in [\theta_{i,j}^{\min}, \theta_{i,j}^{\max}], ~~\forall~ i,j \in \mathscr{N}, 
  \label{const:phase}
\end{equation}
\textcolor{black}{where $s_{i,j,t}$ denote the apparent power flow from node $i$ to $j$ at time $t$.}
\eqref{eq:voltage}, \eqref{const:thermal} and \eqref{const:phase} denote the voltage constraint for nodes, thermal limit constraint and phase angle constraints for branches, respectively. \eqref{eq:powerbalance} denotes the nodal balance of active and reactive power in the network.
\eqref{const:genlimit} denotes the generator output power limits.
\eqref{eq:ohmslaw} denotes Ohm's law. 
\end{subequations}

\textcolor{black}{To avoid that both the ramp up and down flexibility described in \eqref{eq:flexdef} are non-zero simultaneously they have been defined as $\Delta P_{i,t}^{\text{flex}+}\geq 0, \Delta P_{i,t}^{\text{flex}-}\leq 0$. As such a change in either of them increases the objective function value in \eqref{eq:porgobjective}.
Active power of the generation is not included in the objective function, as 
this would influence the amount of flexibility activation. 
Especially, optimization with several components which may come in conflict, it is not clear how these components will affect each other.
Furthermore, the active power dispatch is an entirely different problem. The goal of this study is to find out the flexibility needs of a DN provided active power is already dispatched.}

\subsection{Robust FNA under uncertainty}
\textcolor{black}{Robust optimization is a two-stage optimization problem where the inner \textcolor{black}{level} minimizes the objective function and the outer \textcolor{black}{level} maximizes the inner-optimization within an uncertainty set \cite{conejo2021robust, baringo2010offering, bertsimas2006robust,ben2002robust}.
\textcolor{black}{For the robust FNA, we utilize scenarios for defining the uncertainty set.}
The inner \textit{min}-problem solves the FNA-OPF for all generated scenarios. The outer \textcolor{black}{level} \textit{max}-problem finds the robust flexibility levels for all time steps at a day-ahead level.
The ramp down and ramp up flexibility needs of a DN is modelled as an empirical cumulative distribution function (ECDF), given as
\begin{equation}
    \begin{split}
        F_{i,t,+}(z) & = \frac{1}{S} \sum_{j=1}^S \mathbf{I}_{(-\infty,z]}(\Delta P^{\text{flex}+}_{i,t} \text{for scenario } j), \forall i,t, \\
        F_{i,t,-}(z) & = \frac{1}{S} \sum_{j=1}^S \mathbf{I}_{(-\infty,z]}(\Delta P^{\text{flex}-}_{i,t} \text{for scenario } j),\forall i,t, \\
    \end{split}
\end{equation}
where identity function is given as
\begin{equation}
\mathbf{I}_{(-\infty,z]}(y)=
\begin{cases}
1 ,& \text{if }y\leq z ,\\ 
0 , & \text{otherwise,} 
\end{cases}
\end{equation}
Based on ECDF, the chance constrained robust flexibility needs of the distribution network are identified as below
\begin{equation}
    \begin{split}
        (\Delta P^{\text{flex}+}_{i,t})^* & = F_{i,t,+}^{-1}(1-\epsilon_{cc}), \\
        (\Delta P^{\text{flex}-}_{i,t})^* & = F_{i,t,-}^{-1}(\epsilon_{cc}), \\
    \end{split}
    \label{ccequation}
\end{equation}
where $\epsilon_{cc} \in [0,1]$ denotes the chance constraint level.
}

The load profiles and their associated forecast error profiles are utilized by generating 1000 scenarios. These scenarios are used as input for FNA-OPF. Note that to obtain a feasible solution in all the scenarios, we solve FNA-OPF where each load and/or generation can be curtailed. The activated flexibility is analysed as shown in Fig.~\ref{fig:flexnee}. The chance constraint level will denote the amount of risk in the procurement of flexibility. If the $\epsilon_{cc}$ is too low, then the risk for over-procurement is high. However, if the $\epsilon_{cc}$ is too high, then the risk of under-procurement is high, see \eqref{ccequation}.
Based on a CC level, the nodal ramp up and ramp down power and energy needs of a DN are identified. 
These nodal needs' assessment is aggregated into the zones identified in Section \ref{section3}. 

\pagebreak

\section{Numerical case studies}
\label{section6}

\subsection{System description}
The DN considered is an adaptation of one of the Spanish LV feeders described in \cite{koirala2020non}.
The DN consists of 76 nodes and 75 branches connecting the nodes.
{52 loads are connected in 28 DN nodes, implying many nodes have more than one load connected.}
The network diagram is shown in Fig.~\ref{fig:network} (in Appendix \ref{appendix2}).
\textcolor{black}{The network used in this work along with load profiles can be found on GitHub \cite{githubfna}.
The numerical simulations are performed using PowerModels.jl in Julia / JuMP
\cite{coffrin2018powermodels}.
Since the FNA-OPF problem is nonconvex, therefore, IPOPT solver \cite{wachter2006implementation} is utilized to solve \eqref{eq:fnaoptimization}. The feasibility of the solution of FNA-OPF is verified using power flows.}

\begin{figure}[!htbp]
	\center
	\includegraphics[width=4.5in]{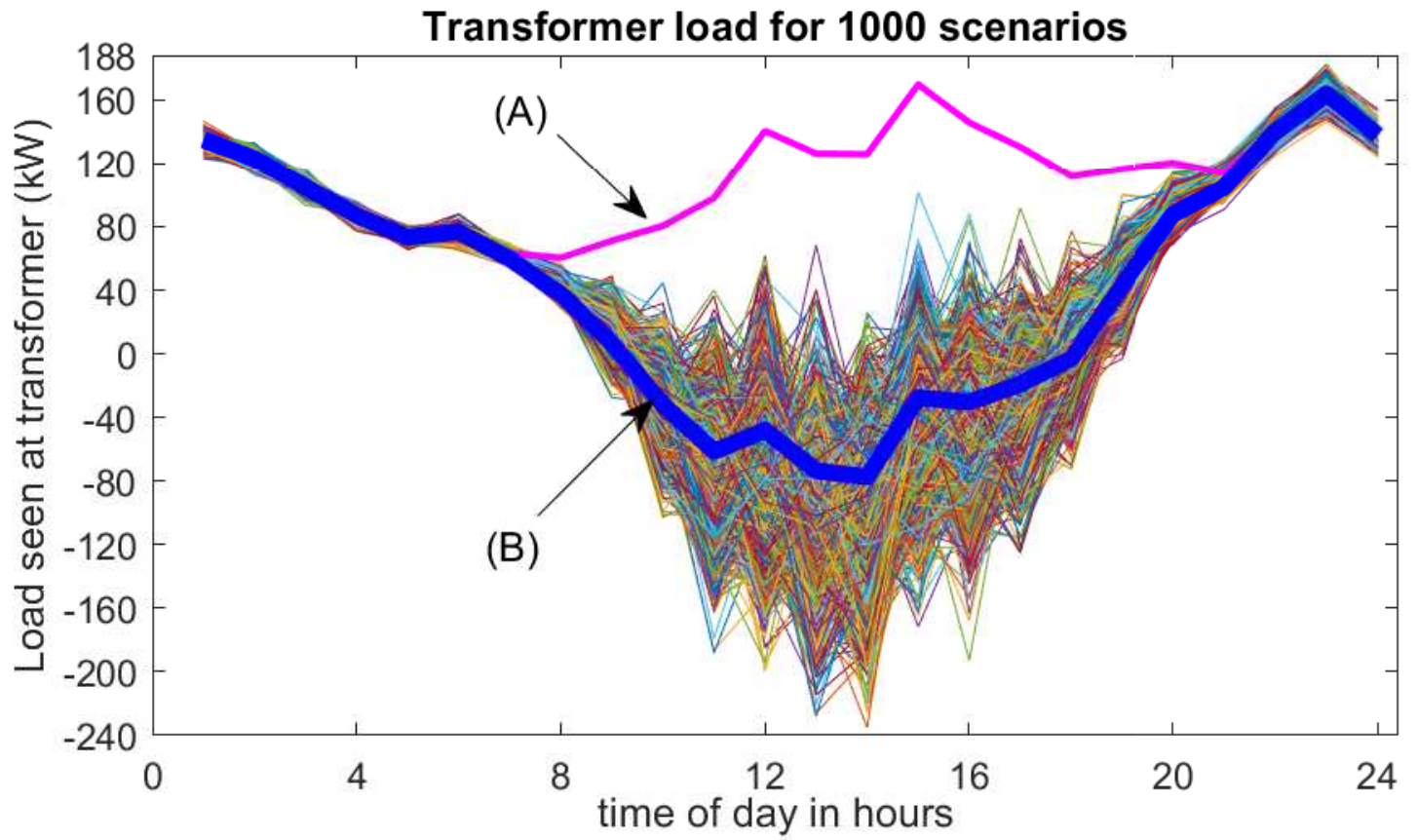}
	\vspace{-6pt}
	\caption{\small{Net load seen at transformer with 1000 scenarios. (A) shows the nominal aggregate load profile without PV generation, (B) shows the nominal aggregate load profile with PV generation.}}
	\label{fig:transload}
\end{figure}

The load and PV distributions are selected so as the network issues are visible for significant periods.
The total load of the DN seen at substation over a day is 2.719 MWh.
Total installed PV is equal to 338 kW peak, with a cumulative PV generation over a day of 1.709 MWh.
Based on the above load profiles, {referred to as nominal nodal profiles}, 1000 scenarios are generated using Cholesky decomposition and shown in Fig.~\ref{fig:transload}.
For the scenario generation, the forecast error for nodal load is assumed to be 30\% and PV generation is assumed to be 40\%.
Note from Fig.~\ref{fig:transload}, that the load perturbations are reduced during aggregation at the substation, however, PV generation due to high simultaneity factor leads to significantly high fluctuations during the day.
For the nominal load profile, the self-sufficiency is 62.85\% which means a substantial part of consumer load is met using PV generation.
The voltage set-point at the slack bus is set at 1.01 per unit.

Due to high PV installations and high peak loads during late evening, we observe DN voltage and thermal constraint violations.
For the 1000 scenarios generated, we observe that only 57.03\% of optimal power flow calculations are feasible for all time steps using an iteration limit of 3000.
We observe under voltage (voltage $<0.95$)) for 6.165\% of the samples in all scenarios, and 3.275\% of samples we observe over-voltage (voltage $>1.05$).
{Thermal overload is calculated by performing OPF with relaxed network constraints, as OPF output for infeasible cases are not reliable.
\textcolor{black}{Power flow calculations are used for all non-feasible OPF solutions.}
We observe thermal over-loads for 0.11\% of \textcolor{black}{samples over all samples, i.e., (number of scenarios)$\times$(number of nodes)$\times$(number of time steps in 1 scenario).}
}

\begin{figure}[!htbp]
	\center
	\includegraphics[width=4.5in]{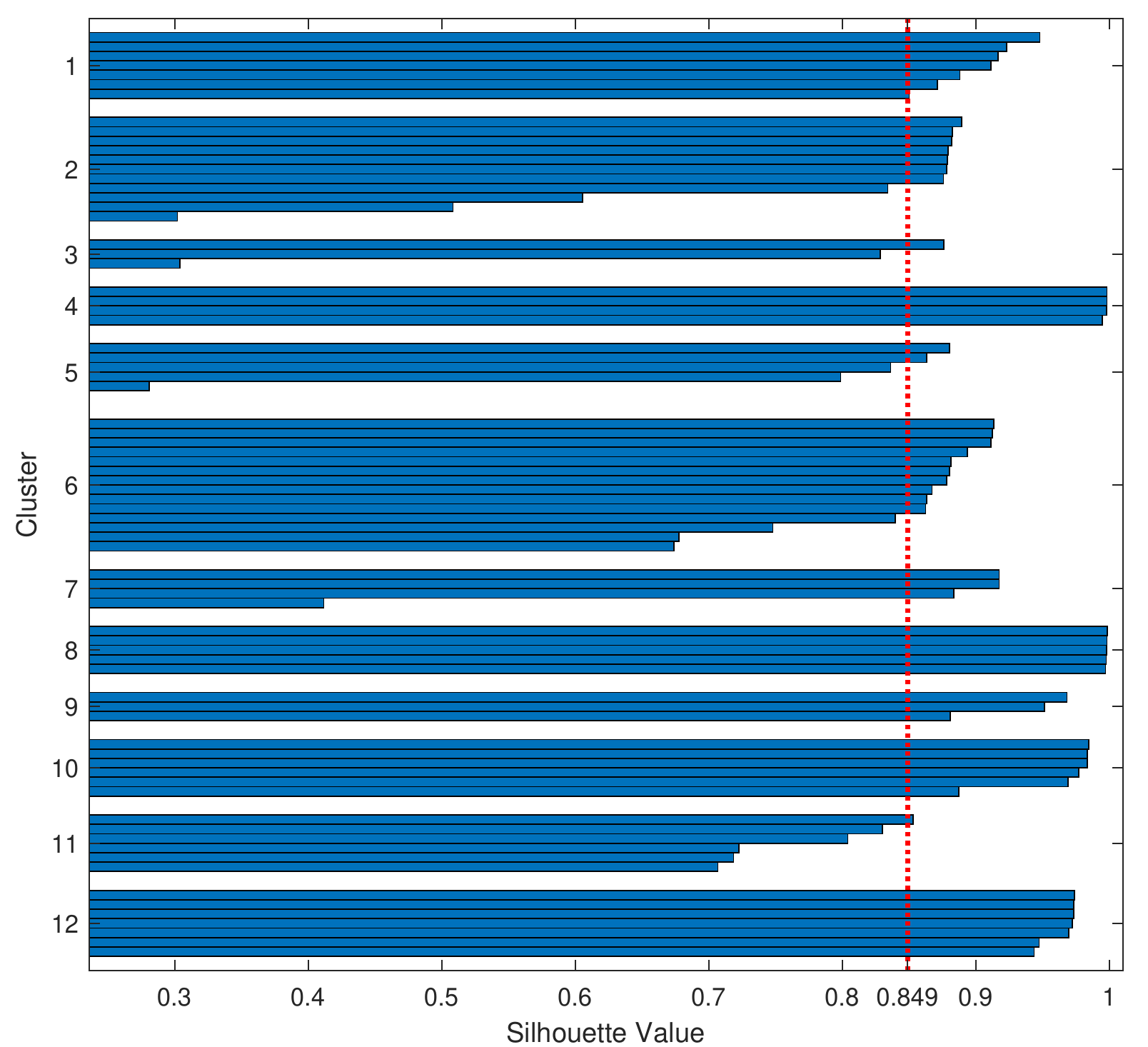}
	\vspace{-6pt}
	\caption{\small{Silhouette score for 76 node network with 12 zones}}
	\label{fig:silhouscore}
\end{figure}

The mean silhouette score for the 76 node network based on the best cluster evaluation described is shown in Fig. \ref{fig:silhouscore}. The mean silhouette score for 12 clusters is 0.8487.
The high silhouette score implies the nodes in a cluster are strongly related to each other and weakly related to other nodes in the other clusters.
Fig.~\ref{fig:network} shows the zones of the DN. 
Note that the naming of zones is arbitrarily selected based on \textit{k}-means centroid.

\vspace{-4pt}
\subsection{Case study 1: Flexibility needs assessment}
\begin{figure}[!htbp]
	\center
	\includegraphics[width=4.5in]{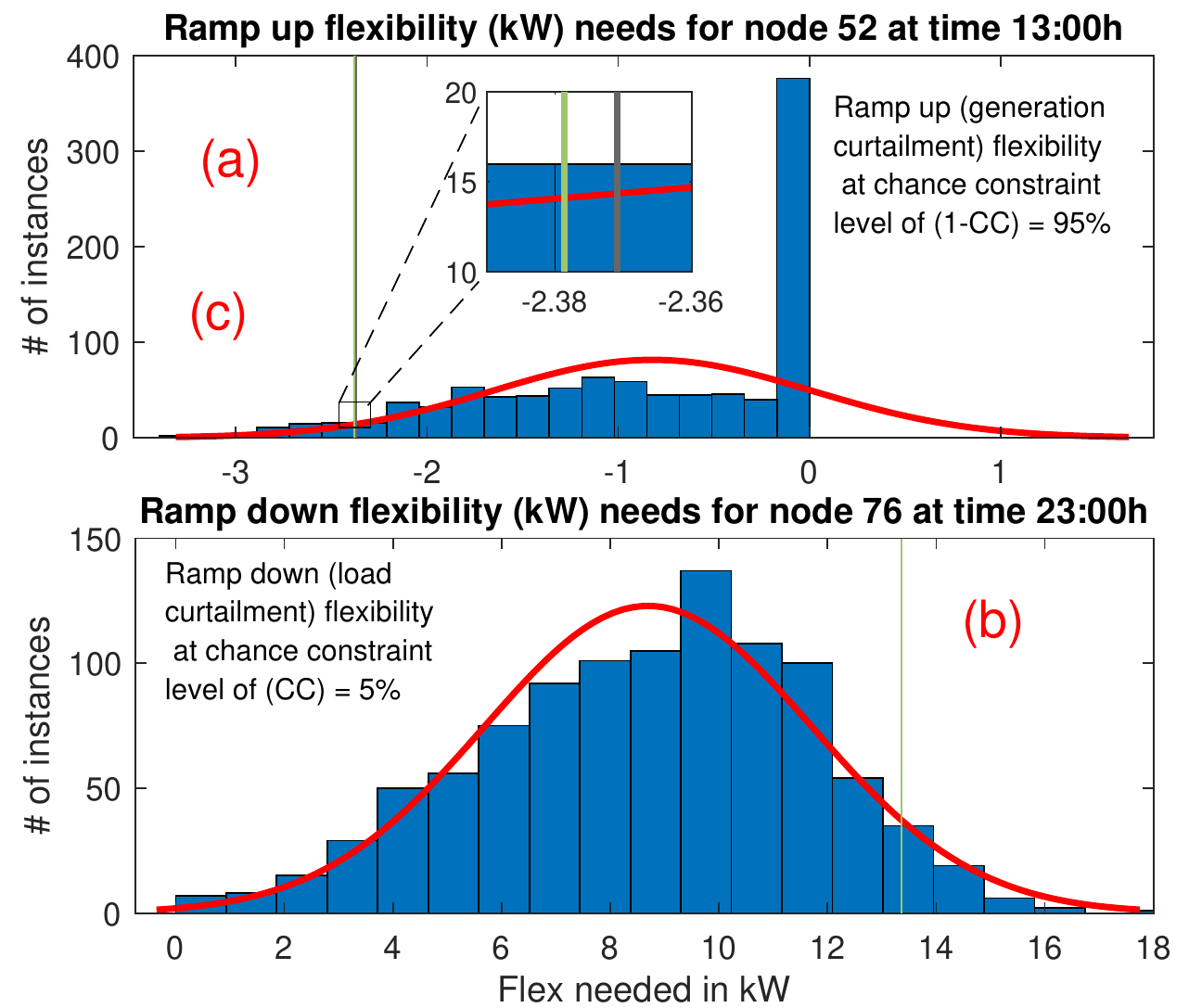}
	\vspace{-5pt}
	\caption{\small{FNA calculation based on normal distribution fit and CC quartiles. \textcolor{black}{The error between inverse ECDF and quantile of approximate normal distribution fit is 0.33\% and 0.02\% respectively.}}}
	\label{fig:flexcalcquartiles}
\end{figure}

FNA-OPF is executed for unbounded power and energy flexibilities. 
Thus, we identify the minimum amount of flexibility needed to avoid all network issues in all scenarios.
FNA based on chance constraint level of 5\% is shown in Figure \ref{fig:flexcalcquartiles}.
The distribution of flexibility needs is fitted using a normal distribution.
Note from Fig. \ref{fig:flexcalcquartiles} that the quality of the approximation  depends on the distribution of the flexibility needs for all scenarios.
\textcolor{black}{We utilize an empirical cumulative distribution function for applying chance constraint as normal distribution fit, shown in Fig. \ref{fig:flexcalcquartiles}(a). From the numerical results we observed that the error between \textcolor{black}{FNA  calculated by applying chance constraint level on} the inverse ECDF and 
on an approximate Gaussian fit is less than 1\% for most points, however, few instances had an error exceeding 5\%. 
Since the distribution of FNA of a DN is not known, ECDF is used
for all subsequent results.}
Note for ramp up flexibility, the chance constraint level is ($1-$ $\epsilon_{cc}$) level because of the sign reversal.

\begin{figure}[!htbp]
	\center
	\includegraphics[width=5in]{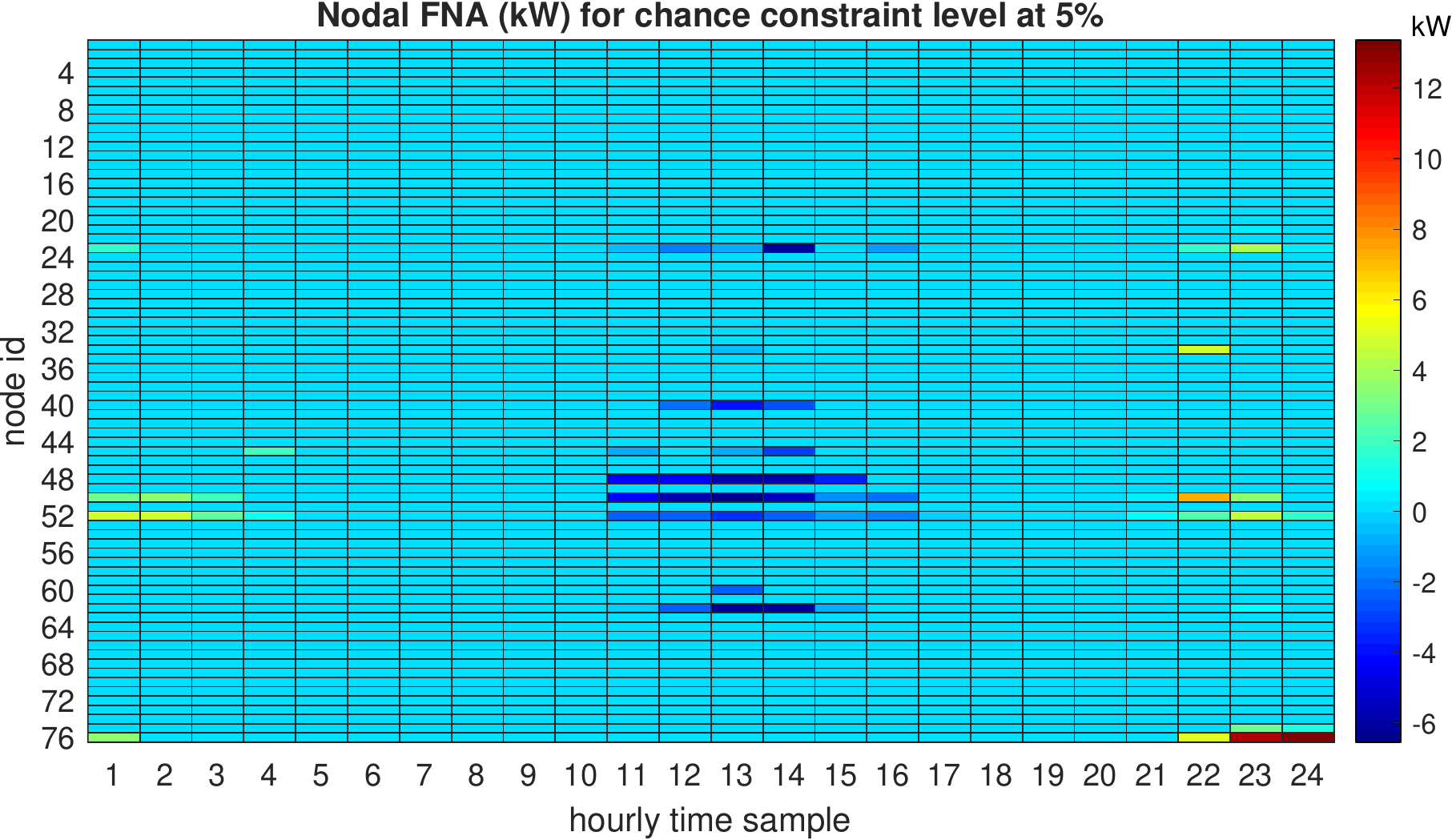}
	\vspace{-5pt}
	\caption{\small{Nodal flexibility power needs with $\epsilon_{cc}=0.05$.}}
	\label{fig:nodalFlexneed}
\end{figure}

\begin{figure}[!htbp]
	\center
	\includegraphics[width=5.0in]{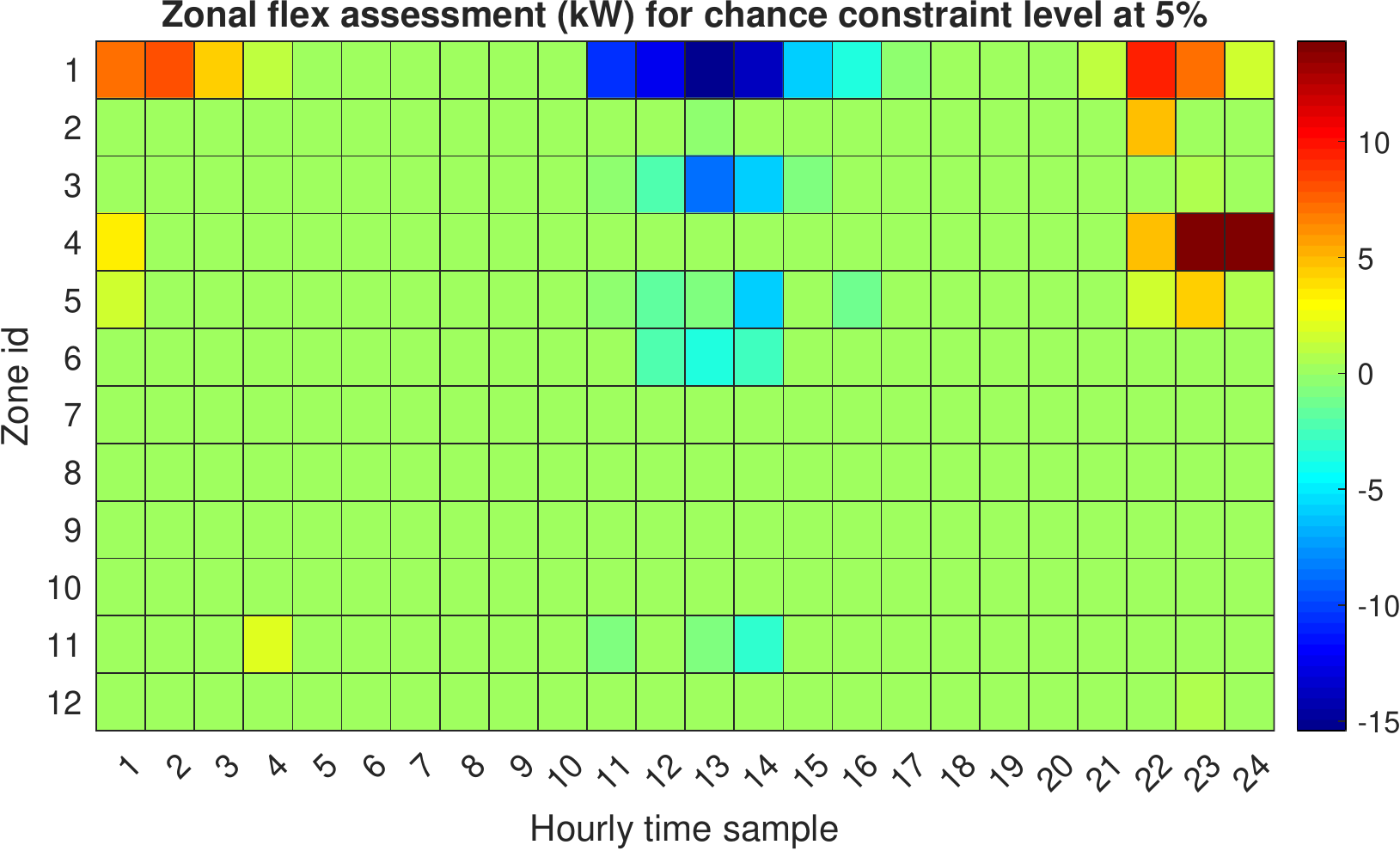}
	\vspace{-5pt}
	\caption{\small{Zonal flexibility power needs with $\epsilon_{cc}=0.05$}}
	\label{fig:zonalFlexneed}
\end{figure}

Fig. \ref{fig:nodalFlexneed} shows the nodal ramp up and ramp down flexibility power needs of the DN.
The nodal needs are aggregated into zones of the DN. Zonal flexibility power needs are shown in Fig. \ref{fig:zonalFlexneed}.
The figure shows that the majority of generation and load flexibility is needed in zones 1, 3, 4, 5, 6, and 11. Note that these zones are located at the end of the distribution feeder. The locational disparity of prosumers
can be reduced by implementing a flexibility market.
The participation of flexible resources in such a market can create an additional revenue stream for such prosumers and can alleviate the locational disparity.

From Table \ref{tab:compareZonalNodal} it can be observed that normalized flexibility needs if aggregated in zones will have a reduction in standard deviation (STD) compared to nodal flexibility needs. The mean and STD denotes the parameters for the normal distribution fitted to the histogram of ramp up and down flexibility power needs.
We observe that in this numerical case study that the \textit{zonal} flexibility needs have an STD \textcolor{black}{which is} $43.48\%$ and $57.85\%$ less for ramp up and down \textit{nodal} flexibility needs.
Thus, zonal flexibility needs assessment is more immune to uncertainty.
\textcolor{black}{
The aggregation of uncertain parameters leads to the reduction in variance of uncertainty is a well known concept \cite{miettinen2020simulating}. 
}


\begin{table}[ht]
 \footnotesize
	\caption {Comparing aggregate nodal and zonal flexibility variation} \vspace{-5pt}
	\label{tab:compareZonalNodal}
	\begin{center}
		\begin{tabular}{c|c|c|c|c}
			\hline
			 & \multicolumn{2}{|c|}{\textbf{Nodal flex need (kWh)} }& \multicolumn{2}{|c|}{\textbf{Zonal flex need (kWh)}} \\
			\hline 
			{Perf. index}& Ramp Up & Ramp Down & Ramp Up & Ramp Down\\
             \hline 
             \hline
            Mean & - 250.39 & 328.7 & - 250.39 & 328.7 \\ 
             \hline
            STD & 328.5 & 256.6 & 185.8 & 108.2\\ 
             \hline
            STD reduction & - & - & \textbf{43.48}\% & \textbf{57.85}\%\\ 
            \hline
        \end{tabular}
		\hfill\
	\end{center}
\end{table}
Table~\ref{tab:compareZonalNodal} shows that a significant amount of ramp up and ramp down flexibility is needed for DN, with network issues less than 10\% of the time.
In our numerical example, mean ramp up and down needs are around 9.2\% and 12.1\% respectively.
With growing DG integration, the need for flexibility planning will be crucial for maintaining the reliability of DN.

\begin{figure}[!htbp]
	\center
	\includegraphics[width=4.5in]{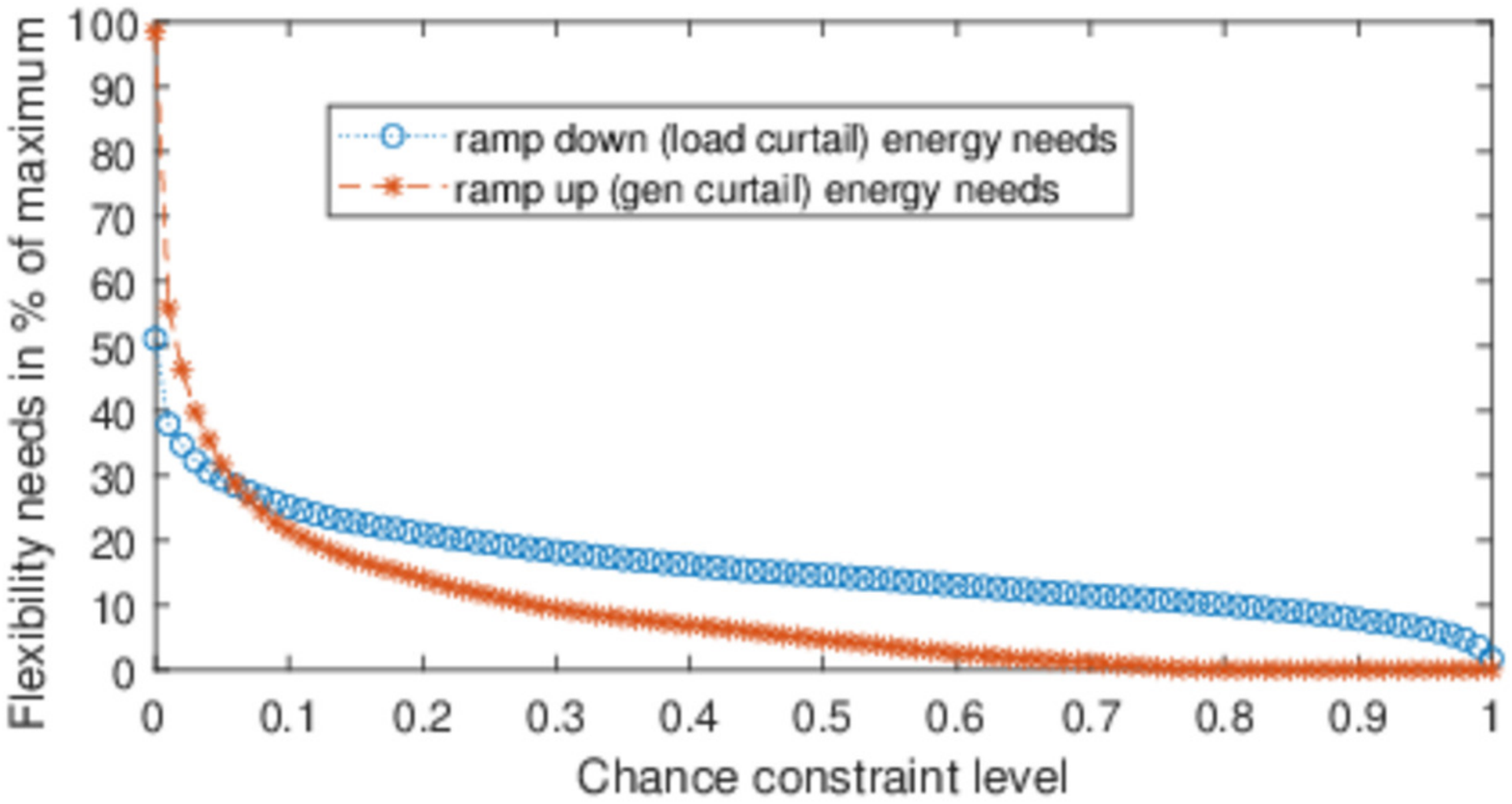}
	\caption{\small{Flexibility needed with chance constraint level}}
	\label{fig:flexNeed}
\end{figure}

Fig. \ref{fig:flexNeed} shows the impact of the chance constraint level on the flexibility energy needs. The heavy tail distribution can be avoided by a small value of CC. Choosing $\epsilon_{cc}=$ 5\% can reduce ramp up energy needs by 68\% and ramp down energy needs 40\%, respectively.
The CC level can also be used for identifying the criticality of a  resource in ensuring DN reliability, as shown in the next section.
\vspace{-5pt}

\subsubsection{Projecting CC levels on network issues}
{The CC levels are analogous to the risk DSOs are willing to take while planning for DN flexibility in a day-ahead setting.
It will be useful to understand how the CC levels project on avoiding probable network issues. We perform traditional OPF calculations with different levels of CC for FNA. The network voltage and thermal congestions are analysed and listed in Table \ref{tab:projectCC}.
where the mean values of probabilities over 1000 scenarios are provided.
Note for the $\epsilon_{cc}=$ 5\%, the mean under-voltage probability reduced from 6.165\% for no flexibility to 0.076\%, a reduction of 98.77\%. The mean over-voltage probability is reduced by 84.2\% and thermal over-load probability is reduced by 77.9\%. The mean hours of congestion (thermal and/or voltage) for $\epsilon_{cc}=$ 5\% is reduced by 93.23\%. 
As expected, a robust FNA with $\epsilon_{cc}=0$ will eliminate all DN issues.
From Tab. \ref{tab:projectCC} we observe that CC levels cannot be directly projected on probable DN issues and will be governed by DN layout and the nodal distribution of demand. Next, we discuss the mechanism for selecting the appropriate CC-level for FNA.
}

\begin{table}
\centering
\caption {Projecting CC levels on probabilities of network congestion} \vspace{-2pt}
	\label{tab:projectCC}
\begin{tabular}{l|l|llll} 
\hline
                                                         & \begin{tabular}[c]{@{}l@{}}CC \\level\\\%\end{tabular} & \multicolumn{1}{l|}{\begin{tabular}[c]{@{}l@{}}Under\\voltage \\\%\end{tabular}} & \multicolumn{1}{l|}{\begin{tabular}[c]{@{}l@{}}Over\\voltage\\\%\end{tabular}} & \multicolumn{1}{l|}{\begin{tabular}[c]{@{}l@{}}Thermal \\over \\load \%\end{tabular}} & \begin{tabular}[c]{@{}l@{}}\% of \\hours~with \\congestion\end{tabular}  \\ 
\hline
\hline
\begin{tabular}[c]{@{}l@{}}No \\flexibility\end{tabular} &                                                     -   & 6.165                                                                            & 3.275                                                                          & 0.110                                                                                 & 46.36                                                                    \\ 
\hline
                                                     & 0                                                      & 0                                                                                & 0                                                                              & 0                                                                                     & 0                                                                        \\
With                                                     & 1   &    0.066&    0.102  &  0.007 &   1.108    \\
robust                                                   & 5                                                      & 0.076                                                                            & 0.518                                                                          & 0.024                                                                                 & 3.138                                                                    \\
CC                                                       & 10  & 0.203  &  0.915 &   0.044 &   6.196                                     \\
FNA                                                      & 20                                                     & 0.544  &  1.440 &   0.082  & 12.16                  \\
planning                                                 & 30                                                     & 0.996&    1.891 &   0.117 &  17.63                                                                       \\
                                                         & 40                                                     & 2.079 &   2.563 &   0.153&   26.43                                                                      \\
\hline
\end{tabular}
\end{table}

\subsubsection{Pareto optimal tuning of CC level}
\begin{figure}[!htbp]
	\centering
	\includegraphics[width=5in]{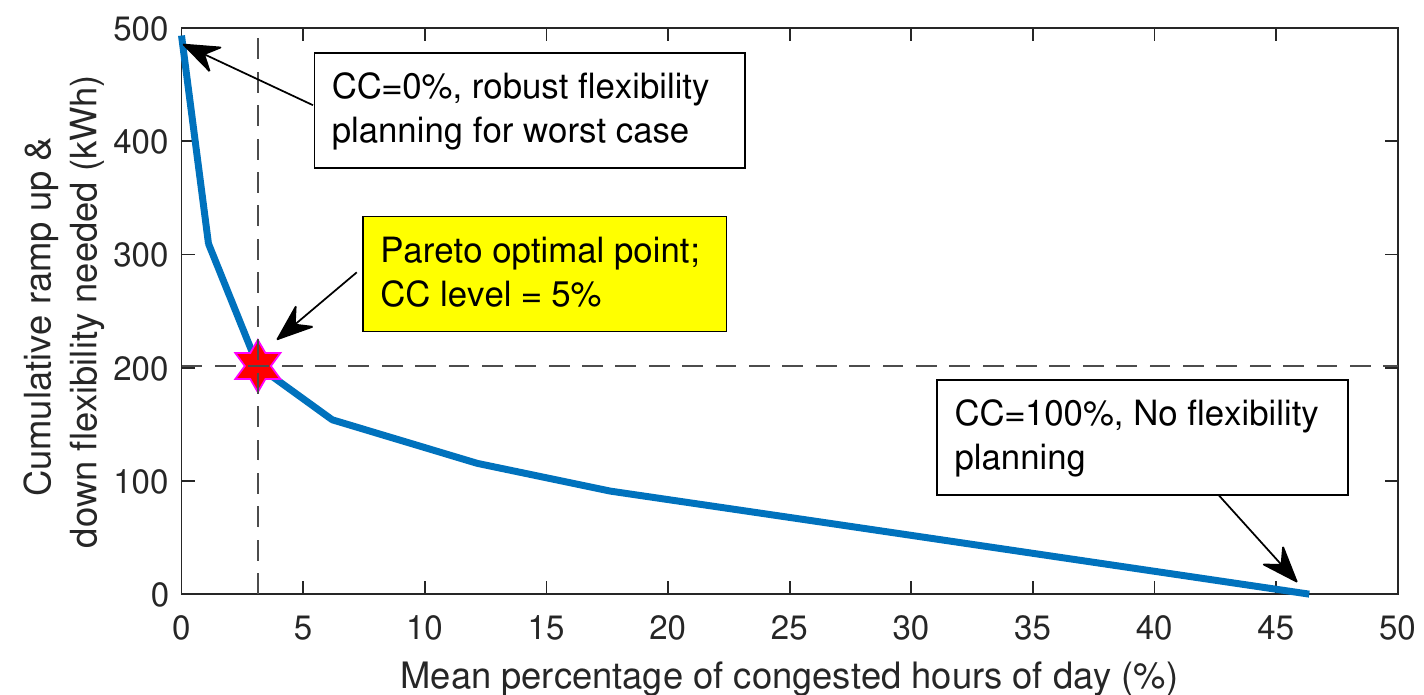}
	\vspace{-7pt}
	\caption{\small{Tuning CC level using Pareto optimality.}}
	\label{fig:pareto}
\end{figure}

The FNA of a DN is governed by network layout and nodal load profiles. 
A Pareto front is built for the two conflicting goals of reducing the amount of flexibility needed, and probable network congestion. Fig. \ref{fig:pareto} shows that the Pareto optimal value of the CC-level for the considered DN is 5\%.

\subsection{Case study 2: Marginal value of energy and power}
\begin{figure*}
	\center
	\includegraphics[width=7in]{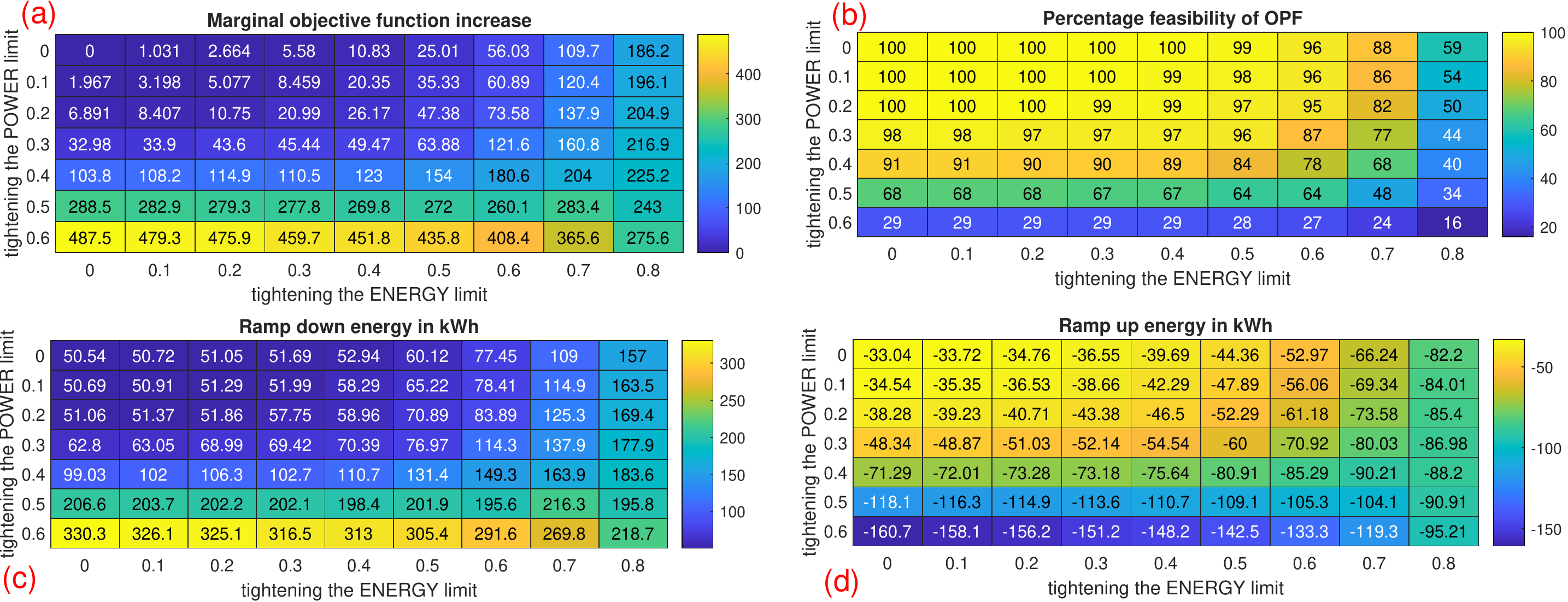}
	\vspace{-5pt}
	\caption{\small{DN flexibility needs with flexibility energy and power bound tightening; (0,0) denotes unconstrained solution of FNA-OPF.}}
	\label{fig:marginal}
\end{figure*}

Flexible resources are valued differently in different types of markets \cite{hashmi2018long}. 
The goal of this case study is to quantify the marginal impact of energy and power constraints for flexible resources. 
The ramp up and down energy and power constraints are tightened in 10\% steps, in order to evaluate:
(a) marginal increase in objective function value of FNA-OPF,
(b) calculate the percentage feasibility of 100 scenarios,
(c) cumulative ramp up and
(d) cumulative ramp down energy needs.
These metrics are shown in Fig. \ref{fig:marginal}.
The key observations from this study are:
\begin{itemize}
    \item The marginal cost for ramp up and down power is more than two times that of the energy. This is observed in Fig. \ref{fig:marginal}(a) where the marginal increase in the objective function is more than 487 \% compared to the unconstrained objective function value for FNA-OPF. This marginal increase is due to a reduction in flexibility power ranges by 60\%. However, for up to 80\% reduction in energy constraint over a day, the marginal increase in the objective function is around 186\%. Similar trends can be observed for other combinations of simultaneous tightening of power and energy constraints.
    \item The percentage of feasible FNA-OPF calculations is shown in Fig. \ref{fig:marginal}(b). It is expected that the feasibility will reduce with tighter flexibility power and energy constraints.
    Note that we limit the levels of power and energy bounds tightening to 0.6 and 0.8 respectively, as the majority of points are infeasible for any further tightening. The solutions with extremely high levels of tightening are unreliable, as in such a case either there is no solution possible or a maximum number of iterations is reached. In both these cases, the solution for FNA-OPF cannot be used.
    \item Fig. \ref{fig:marginal}(b) also indicates the combinatorial nature of the flexibility needs assessment problem. We can observe that there are many combinations of feasible solutions that can solve the same network congestion.
    \item The ramp up and ramp down energy are shown in Fig. \ref{fig:marginal}(c) and (d). It is observed that as power and energy constraints are tightened, an larger amounts of flexibility is needed somewhere else in the network to mitigate solve congestion. As the distance from the congestion location increases, the efficiency of the procured flexibility to solve congestion decreases. 
\end{itemize}

\section{Conclusion}
\label{section7}
Power system flexibility will be crucial for reliable operation of distribution network for ensuring {a large scale integration of} 
distributed generation and new loads such as electric vehicle.
An optimization-based framework for calculating the day-ahead flexibility needs of a distribution network is presented in this work.
This flexibility needs assessment tool considers future uncertainties in the form of PV generation and demand scenarios.
The FNA-OPF identifies nodal ramp up and ramp down power and energy needs of a distribution network, which minimizes the cost of dispatching flexible resources.
Based on the distribution of flexibility needs, a chance constraint-based robust needs assessment is performed to avoid under or over procurement of such resources.
In the first case study, this needs assessment is evaluated for a zonal partition of the distribution network. The \textcolor{black}{appropriate} number of zones is determined based on electrical distance measure and spatial partitioning.
It is observed that zonal needs assessment is more immune to future uncertainties, and DSOs can utilize this feature along with nodal flex needs assessment for flexibility planning and operation.
The second case study illustrates the impact of the flexibility bounds for power and energy on the flexibility needs and the marginal cost of additional flexibility. We observe that the marginal impact of tighter power flexibility constraints on activation cost is more than twice compared to energy  flexibility constraint.
{Finally, a mechanism for DSOs to tune the level of chance constraint for FNA-OPF, considering the flexibility needed and probable network congestion reduction, is demonstrated in this paper.}

In future work, we use the need assessment for developing a strategy to select bids in flexibility markets.
The purpose of zonal analysis is to assist DSOs in finding alternative resources in the same zone in case the nodal flexibility needs identified are not available in the flexibility market.
Further, the impact analysis on power and energy flexibility constraints can be utilized to determine the best suitable location and size of flexibility sources in the network. 


\bibliographystyle{IEEEtran}
\bibliography{reference}

\pagebreak

\appendix

\section{Zone selection algorithm}
\label{zoneselection}
The silhouette coefficient of node $j \in M_x$ is calculated as,
$
    s_j = \frac{b_j - a_j}{\max \{b_j, a_j\}},
$
where $a_i$ denotes the mean distance between node $j$ and all other nodes in the group $M_x$ and $b_j$ denotes the mean distance between node $j$ and all other nodes not in the group $M_x$.
The mean silhouette score of a group $M_x$ is given as
$
    S_x = \frac{1}{|M_x|} \sum_{i=1}^{|M_x|} s_i,
$
where $|M_x|$ denotes the number of nodes in the group $M_x$.
The mean silhouette score of a $k$ partitioned network is given as
$
    \text{SC}_T^k = \frac{1}{k} \sum_{x=1}^{k} S_x.
$
For a given network, vary $k$ to maximize the value of $\text{SC}_T^k$. Note $\text{SC}_T^k$ may be high for a small number of clusters which may not suit the application, therefore, the selection of best $k$ which increases $\text{SC}_T^k$, is a design problem.

The algorithm for selecting the best number of zones that maximizes the mean silhouette score is detailed below.
The output of the algorithm is the ideal number of zones for our flexibility assessment application.
\begin{algorithm}
	\small{\textbf{Inputs}: Network details, $\text{SC}_T^k=[~]$,
	\begin{algorithmic}[1]
	    \State Calculate admittance matrix, $Y$, for the network,
	    \State Make the electrical distance matrix double stochastic,
	    \State Set value of $k=2$ clusters.
	    \State Calculate $k$ the largest eigenvalues and eigenvectors.
	    \State Use $k$-means clustering and calculate $\text{SC}_T^k$ and concatenate,
	    \State Increment $k$ till $k\leq N$ and Goto Step 4,
	    \State Analyse $\text{SC}_T^k$ vector to select the best suited number of zones. {The best suited number of zones depends on (a) how many clusters will meet the required application for which clusters are formed, and (b) silhouette score of the cluster. For example, if the silhouette score is very high for very few numbers of clusters, but it does not meet our purpose, then we may opt for a lower silhouette score. }.
	\end{algorithmic}
	\caption{\texttt{Zones of distribution network}}}
	\label{alg:zones}
\end{algorithm}

\section{Network diagram}
\label{appendix2}
\begin{figure*}[!htbp]
	\center
	\includegraphics[width=6.4in]{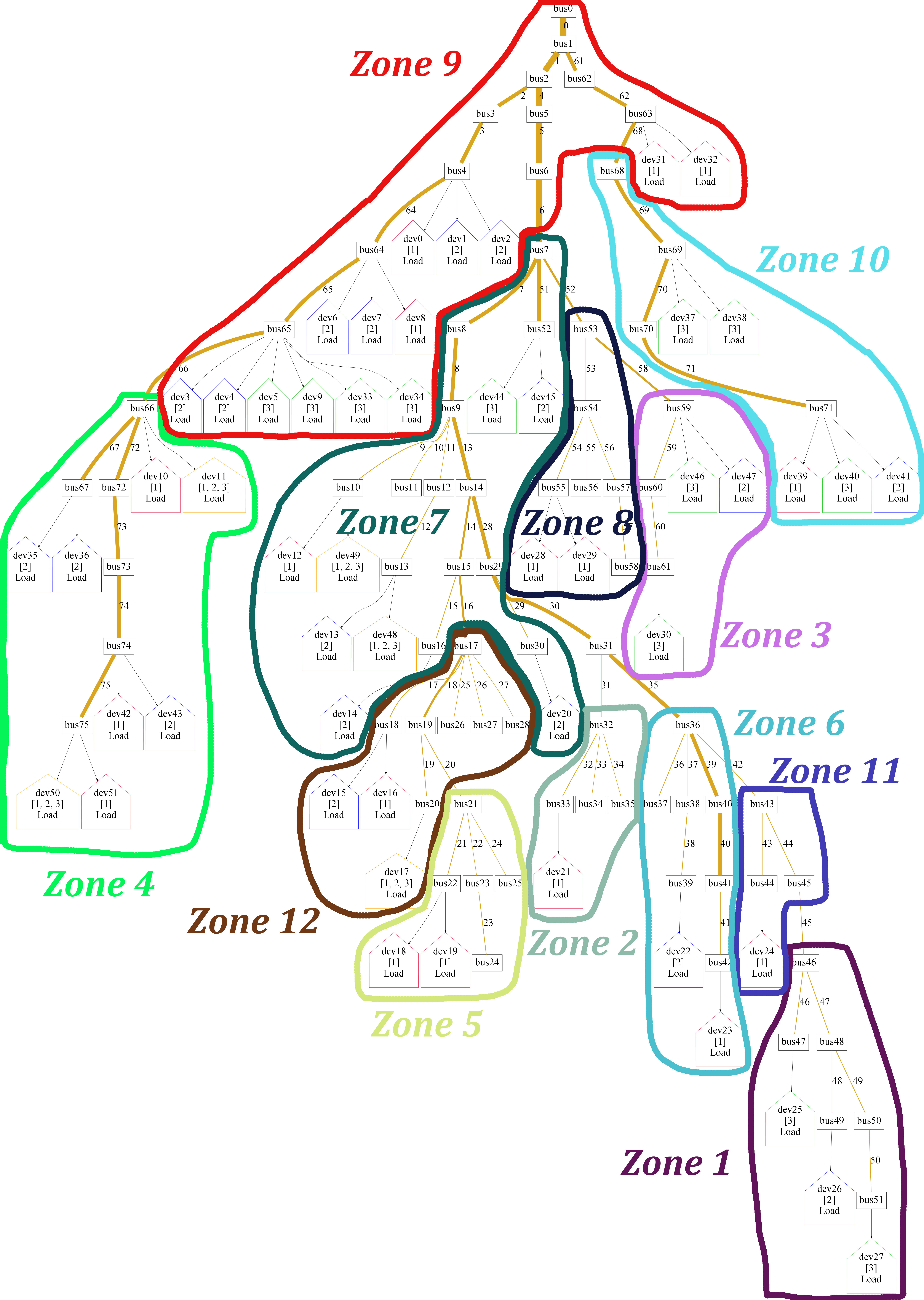}
	\vspace{-10pt}
	\caption{\small{Network diagram with 12 zones indicated}}
	\label{fig:network}
\end{figure*}

The network diagram and the zones are shown in Fig.~\ref{fig:network}.

\end{document}